\documentclass[a4paper,fleqn,usenatbib]{mnras}

\usepackage{amsmath}
\usepackage{amsfonts}
\usepackage{amsbsy}
\usepackage{amssymb}
\usepackage{amsbsy}
\usepackage[T1]{fontenc}
\usepackage{ae,aecompl}
\usepackage{graphicx}
\usepackage{gensymb}
\usepackage{float}
\usepackage[utf8]{inputenc}
\usepackage{tensor}
\usepackage{bm}
\usepackage{enumerate}

\DeclareMathAlphabet\mathbfcal{OMS}{cmsy}{b}{n}

\defcitealias{riols17c}{RL18a}
\defcitealias{riols18}{RL18b}
\title[]{Dust dynamics  and vertical settling in gravitoturbulent protoplanetary discs}
\author[]{
A. Riols $^{1}$, B. Roux $^{1}$, H. Latter $^{2}$,  G. Lesur $^{1}$
\\
$^{1}$Univ. Grenoble Alpes, CNRS, Institut de Planétologie et d’Astrophysique de Grenoble (IPAG), F-38000, Grenoble, France\\
$^{2}$DAMTP, University of Cambridge, Centre for Mathematical Sciences,
Wilberforce Road, Cambridge CB3 0WA, UK.}

\date{Accepted XXX. Received YYY; in original form ZZZ}

\pubyear{2020}

\begin{document}
\label{firstpage}
\pagerange{\pageref{firstpage}--\pageref{lastpage}}
\maketitle

% Abstract of the paper
\begin{abstract}
Gravitational instability (GI) controls the dynamics of
young massive protoplanetary discs. 
Apart from facilitating gas accretion on to the central
protostar, it must also impact on the process of planet formation: directly
through fragmentation, and indirectly through the turbulent
concentration of small solids.  
To understand the latter process, it is essential to determine
the dust dynamics in such a turbulent flow. For that purpose,
 we conduct a series of 3D shearing box simulations of coupled gas and dust, including
 the gas’s self-gravity and scanning a range of Stokes
 numbers, from $10^{-3}$ to $\sim 0.2$.  First, we show that the vertical
 settling of dust in the midplane is significantly impeded by gravitoturbulence, with
 the dust scale-height roughly 0.6 times the gas scale height for centimetre grains. This is 
a result of the strong vertical diffusion issuing from (a) small-scale
 inertial-wave turbulence feeding off the GI spiral waves and (b) 
the larger-scale vertical circulations that naturally accompany the spirals.
  Second, we show that at $R=50$ AU concentration events involving sub-metre
 particles and yielding order 1 dust to gas ratios
 are rare and last for less than an orbit. Moreover,
 dust concentration is less efficient in 3D than in 2D
 simulations. We conclude that GI is not especially prone
to the turbulent accumulation of dust grains.
Finally, the large dust scale-height measured in
 simulations could be, in the future,
 compared with that of edge-on discs seen by ALMA, 
thus aiding detection and characterisation of GI in real systems. 
\end{abstract}

\begin{keywords}
turbulence --- instabilities --- dust ---
protoplanetary discs   
\end{keywords}

%%%%%%%%%%%%%%%%%%%%%%%%%%%%%%%%%%%%%%%%%%%%%%%%%%

%%%%%%%%%%%%%%%%% BODY OF PAPER %%%%%%%%%%%%%%%%%%

\section{Introduction}
\label{sec_intro}
Gravitational instability (GI) manifests within (almost) the entire spectrum 
of astrophysical discs: from
planetary rings and young
protoplanetary (PP) discs, to active galactic nuclei (AGN) and spiral
galaxies. 
It redistributes angular momentum, thus enabling accretion (both
steady and bursty); it generates large-scale structure in the form of
dramatic spiral waves; and it regulates the fragmentation of the disk 
 into bound objects such
 as planets (or stars).  
The critical parameter governing the onset of GI
is the Toomre $Q$ \citep{toomre64},  
\begin{equation}
Q=\dfrac{c_s\kappa}{\pi G \Sigma_0},
\label{mass_eq}
\end{equation} 
where $c_s$ is the sound speed, $\kappa$ the epicyclic frequency,
and $\Sigma_0$ the background surface density. 
In a razor thin disk, linear axisymmetric
disturbances are unstable when $Q<1$, though
\emph{nonlinear} non-axisymmetric instability can occur for a critical $Q\gtrsim
1$. In PP disks, this criterion translates to $M_\text{disk}\gtrsim 0.1 M_\text{star}$, where
$M_\text{disk}$ and $M_\text{star}$ are the masses of the disk and central star.
 Depending on the speed of the cooling process, the instability either
forces the disk to fragment
 or saturates in a gravito-turbulent state characterised by spiral density
 waves \citep{Gammie2001, rice03, rice06,durisen07}. 

Indeed, large-scale
`grand-design' spirals have been observed in several PP disks (e.g.\
 {Elias 2-27, WaOph 6, MWC758) and more disordered `streamers' 
in FU Ori systems, structures that might be attributable to GI
\citep{liu16,dong16,perez16,huang18}}. But it should be emphasised that 
only very massive, and thus very unstable, disks ($M_\text{disk}> 0.25 M_\text{star}$)
generate observable structure:
spirals associated with more moderate gravitoturbulence may be too
flocculent to be detected with current facilities \citep[e.g.][]{dong15}.
On the other hand, the presence or not of GI can be inferred 
from calculations of $M_\text{disk}$: recent surveys find that 50\%
of class 0 and 10-20\% of class I sources might be unstable to GI 
\citep{tobin13, mann15}, though such estimates are problematised
by the difficulty in reliably determining these disk masses.

It has been pointed out that  {Class II} and older disks possess
masses
that are too small in comparison to those of observed exoplanetary
system,
a fact that has tempted researchers to conclude that planets form early
(and/or most disk accretion occurs early)\citep{najita14,manara18}. 
This idea is reinforced by the prevalence of ring
structure
in young disks (e.g.\ HL Tau and GY 91), which are generally thought
to be caused by speedily formed planets \citep{alma15,sheehan18}. 
Taken together, these points put forward a case that GI
is operating precisely when planet formation is active. It thus motivates us
to look into the role (if any) GI assumes during early planet formation in PP disks.
 
A first step is to establish the dynamics of intermediate size ($\mu$m to
m) dust grains when aerodynamically
coupled to the gravitoturbulent gas. 
In fact, a series of studies in 2D discs
\citep{gibbons12,gibbons15,shi16} reveal that GI spiral waves
can entrain and aggregate dust particles, thus
facilitating their growth through the difficult mm to m size range, in
which various barriers halt their growth. 
By enhancing their densities, such aggregates may induce streaming
instability  (when $\rho_d \gtrsim \rho_g$) and/or 
gravitational collapse
\citep[e.g.][]{youdin05,cuzzi08,bai10,shi13,simon15,yang17}. It is not guaranteed,
however,
that this aggregation works as well in 3D stratified discs. 
Of particular concern
are additional vertical flows that may hinder dust sedimentation and/or the accumulation of dust in spirals. 
Very strong spiral shocks induce hydraulic jumps and accompanying fountain flows \citep{boley05}
but, in fact, (less violent) vertical flows accompany spiral waves generically:
recent high-resolution
simulations by \citet{riols17b} and \citet{riols18} demonstrated that GI spiral waves (a) are subject to parasitic
instabilities that produce small-scale inertial-wave
 turbulence, and (b) induce coherent large-scale vertical circulations mediated by g-modes. Both flows are necessarily
absent in 2D simulations, and also potentially difficult to describe in global 3D simulations. 
Nonetheless, they should critically influence
the dynamics of dust. Assessing the impact of these two types of flow is the main goal of this paper.

Quite apart from planet formation, characterising grain
sedimentation may bring new constraints on
observed disc properties and aid detection of GI
in some discs. It is possible with ALMA to directly measure
the size of the dust layer from the continuum sub-millimetre
emission of structured discs \citep[e.g. HL Tau, see ][]{pinte16} or
edge-on discs \citep[e.g. HH30 and many others,
see][]{louvet18,duchene19}. A direct comparison of this size with that
measured in simulations could provide precious information on the
origin and nature of disk turbulence \citep{riols18c}. Because GI can
develop strong supersonic motions, it is expected that the settling
process differs significantly from other type of turbulence
\citep[driven by the magneto-rotational instability or the vertical
shear instability for example, 
see][]{fromang06b,picogna18} and could leave
a detectable imprint on the vertical dust distribution. GI's
resistance
to settling, however, can only be assessed in 3D simulations of the type
we present here. 
  
Another important question concerns the observable properties of the
spiral waves that GI triggers. While many spiral arms have been
observed in various PP discs, some, such as
HL Tau,  are sufficiently massive to be GI  unstable  {\citep{booth20}} and yet
do not show up spiral structures.
One solution to this particular case
is to claim that the GI is not strong enough to generate
detectable `grand-design' spiral structure (see earlier). But it is also possible
that the GI dust structure {(in particular those made of millimeter dust particles traced by instruments like ALMA)}
differs significantly from the GI gas structure. 
 One way to decide on this issue is to understand the relationship
 between characteristic GI features in the dust and in the gas. 
Because of 3D vertical motions associated with the GI, it is likely
that the dust will at best exhibit a `blurred' analogue of large-scale
gas structure.

Our aim in this paper is to revise previous 2D simulations, which cannot describe the
secondary vertical flows exhibited by GI, and global 3D simulations, which usually
cannot afford the resolution to do so. 
For that purpose,  we performed 3D shearing
box simulations of stratified discs including both self-gravity and
dust,  using a modified version of the  PLUTO code.  The dust
population is approximated as a pressure-less multi-fluid made of
different particle sizes, from a few hundreds of micrometres to decimetre. 
The back reaction of the dust on the gas is taken into account, but not the self-gravity of 
the dust itself. As a preliminary step, we use a very simple cooling law of Newtonian form 
and neglect dust coagulation or fragmentation. 
{Note that simulations by \citet{shi13,baehr19} also explored  dust dynamics in 3D self-gravitating discs, but they did so in the fragmentation, not gravitoturbulent,
regime. 
In particular \citet{baehr19} found that dust is efficiently collected into fragments and ultimately collapse to form planetary cores. }

Our main result is that GI turbulent flows powerfully resist the vertical settling of intermediate size particle  (mm to dm): the
quasi-steady dust layers we find possess scale heights comparable to the gas scaleheight $H_g$.  
Motions associated with both large-scale roll motions and small-scale inertial wave turbulence contribute to the vertical diffusion of solids. Another important result is that for the largest particle size probed (Stokes number of $\sim 0.16$), the dust does concentrate into thin filaments (as in 2D) but with a dust to gas ratio $\rho_d/\rho_g$ that barely exceeds 1; three-dimensional vertical motions tend to inhibit concentration. 
Finally, in the horizontal plane, although most of the grains are trapped into spiral waves, 
the dust structures tend to be less sharp and more smeared out than in 2D.  

The paper is organised as follows: in Section \ref{sec_model}, we describe the model and review the main characteristics of dust-gas interaction.  We also present the numerical methods used to simulate the dust-gas dynamics. In Section \ref{sec_withdust}, we first characterize the main properties of gravito-turbulent discs (without the dust component) and explain how we initialize the simulations with dust.  We then calculate the steady-state dust scaleheights, as a function of Stokes number, and quantify the combined effect of small-scale wave turbulence and vertical circulation in grain lofting.  
We finally characterize the horizontal dust grain dynamics associated with GI spiral waves motions, with an eye to the competition between their horizontal ‘smearing out’ and their entrainment in spirals. We conclude in Section \ref{sec_conclusions} by discussing the applications of our work on protoplanetary discs observations and planet formation.

\section{Model and numerical setup}
\label{sec_model}
\subsection{{Governing equations}}
To simulate gas and dust in gravito-turbulent flow, we use the local Cartesian model of an accretion disc \citep[the
shearing sheet;][]{goldreich65,latter17} where the differential rotation is approximated locally
by a linear shear flow  {$-S x\mathbf{e}_y$} and a uniform rotation rate $\boldsymbol{\Omega}=\Omega \, \mathbf{e}_z$, with $S=(3/2)\,\Omega$ for a Keplerian equilibrium. We denote by $(x,y,z)$ the radial, azimuthal, and vertical directions.  We refer to the $(x,z)$ projections of vector fields as their `poloidal components’ and to the $y$ component as their `toroidal’ one. We assume that the gas is ideal, its pressure $P$ and density $\rho$ related by $\gamma P=\rho c_s(T)^2$, where $c_s(T)$ is the
sound speed (allowed to vary) and $\gamma$ the ratio of specific heats. In this paper,
we neglect molecular viscosity. We adopt a multi-fluid approximation in which the gas and the dust interact and exchange momentum through drag forces.
   
The evolution of gas density ${\rho}$, total velocity {field} $\mathbf{v}$ and pressure $P$ obeys
\begin{equation}
\dfrac{\partial \rho}{\partial t}+\nabla\cdot \left(\rho \mathbf{v}\right)=0
\label{mass_eq}
\end{equation}
\begin{equation}
\frac{\partial{\mathbf{v}}}{\partial{t}}+\mathbf{v}\cdot\mathbf{\nabla
  v} +2\boldsymbol{\Omega}\times\mathbf{v} =- \mathbf{\nabla} \Phi
  -\frac{1}{\rho}\mathbf{\nabla}{P}+ {\bm{\gamma}_{d}}, 
\label{ns_eq}
\end{equation}
\begin{equation}
\dfrac{\partial P}{\partial t}+\nabla\cdot (P\mathbf{v})
  =-P(\gamma-1)\nabla\cdot\mathbf{v}-\dfrac{P}{\tau_c},
\label{heat_eq}
\end{equation}
where the total velocity field can be decomposed into a mean shear and a perturbation $\mathbf{u}$: 
\begin{equation}
\mathbf{v}=-S x\,  \mathbf{e}_y+\mathbf{u}.
\end{equation}
$\Phi$ is  the sum of the tidal gravitational potential induced by the central object in the local frame $\Phi_c=\frac{1}{2}\Omega^2 z^2-\frac{3}{2}\Omega^2\,x^2$  and the gravitational potential $\Phi_s$ induced by the disc itself, which obeys the Poisson equation: 
\begin{equation}
\mathbf{\nabla}^2\Phi_s = 4\pi G\rho.
\label{poisson_eq}
\end{equation}
The last term in the momentum equation (\ref{ns_eq}) represents the acceleration $\bm{\gamma}_{d}$ exerted by the dust’s drag force on the gas  (detailed below). The cooling in the internal energy equation (\ref{heat_eq}) is a linear function of $P$ with a typical timescale $\tau_c$ referred to as the `cooling time'. This prescription is not especially realistic but allows us to simplify the problem as much as possible.  We also neglect thermal conductivity and magneto-hydrodynamical effects.  {Note that we do not include heating from stellar irradiation, which can impact the fragmentation threshold \citep{rice11}}\\

The dust is composed of a mixture of different species,  characterizing different grain sizes. Each species, labelled by a subscript $k$, is described by a pressure-less fluid,  with a given density $\rho_{d_k}$ and velocity $\mathbf{v}_{d_k}$. The equations of motion for each species are: 
\begin{equation}
\dfrac{\partial \rho_{d_k}}{\partial t}+\nabla\cdot \left(\rho_{d_k} \mathbf{v}_{d_k}\right)=0,
\label{eq_mass_dust}
\end{equation}
\begin{equation}
\dfrac{\partial \mathbf{v}_{d_k}}{\partial{t}}+\mathbf{v}_{d_k}\cdot\mathbf{\nabla  v}_{d_k} +2\boldsymbol{\Omega}\times\mathbf{v}_{d_k} =-\nabla{\Phi}+ \bm{\gamma}_{g_k},
\label{ns_eq_dust}
\end{equation}
with $\bm{\gamma}_{g_k}$ the drag acceleration imposed by the the gas
on a dust of type $k$.  The term in the gas momentum equation (\ref{ns_eq}) is obtained by conservation of total momentum:
\begin{equation}
\bm{\gamma}_{d}= -\dfrac{1}{\rho} \sum_k \rho_{d_k} \,\bm{\gamma}_{g_k}.
\end{equation}
The drag acceleration acting on particles of type $k$ is given by:
\begin{equation}
 \bm{\gamma}_{g_k}=\dfrac{1}{\tau_s^k}  (\mathbf{v}-\mathbf{v}_{d_k}).
\end{equation}
where $\tau_s^k$ is the stopping time,  a direct  measure of the coupling between dust particles and gas. In this study we assume 
that dust particles are spherical and sufficiently small that they are in the Epstein regime \citep{weiden77}. 
For particles of radius $a_k$ and internal density $\rho_s$ (which should not be confused with the gas or dust densities), the stopping time $\tau_s^k$ is
\begin{equation}
\label{eq_stoptime}
\tau_s^k  = \dfrac{\rho_s a_k}{\rho c_s}.
\end{equation}
 A useful dimensionless quantity to parametrize this coupling is the
 Stokes number of the $k$th dust species
\begin{equation}
\text{St}_k=\Omega \tau_s^k.
\end{equation}
In what follows, for notational ease and because the meaning will always be clear, we will drop the
subscript $k$ and simply refer to the `St of a given species'. Also,
if not stated otherwise, St denotes the Stokes number in the midplane. We note that the effective Stokes number in the disc atmosphere is larger than St, since it is inversely proportional to the density in a stratified medium.

\subsection{Stokes number and particle size}
\label{conversion}
In this paper, we preferentially use the Stokes number rather than particle size to describe the dust dynamics, since St  is a dimensionless quantity which does not depend on the disc properties and geometry.  Nevertheless, to make possible comparison with observed systems, it is helpful to associate the Stokes number to a grain size. 

In the case of a self-gravitating discs with $Q\sim 1$, hydrostatic equilibrium dictates that the surface density 
\begin{equation}
\Sigma \sim  \rho_0  H_g \sqrt{2\pi} \sim   \dfrac{c_s \Omega}{\pi G},
\end{equation}
where $\rho_0$ is the midplane density and $H_g  \lesssim H $ is the self-gravitating disc scaleheight. $H$ is the standard hydrostatic disc scale height $c_{s_0}/\Omega$ with $c_{s_0}$ the sound speed in the midplane of a hydrostatic disc in the limit $Q\rightarrow \infty$. Thus, combining these different relations, and noting $\Omega^2 = G M^\star/R^3$, we obtain:
\begin{equation}
\text{St} \simeq a  \left( \dfrac{\rho_s \sqrt{2\pi}\pi R^2}{2M^\star}\right) (H/R)^{-1}.
\end{equation}
The factor 1/2 comes from a rough estimate of $H_g\simeq H/2$  based on self-gravitating equilibria \citep[see for instance Appendix A of][]{riols17b}.

Next we assume that $\rho_s=2.5$  {g.cm$^{-3}$}, the central object possesses a mass equal to that of the Sun, and the disc 
aspect ratio of 0.1. These assumptions present us the following conversion 
\begin{equation}
\text{St} \simeq 0.028 \,\,  \left(\dfrac{a}{\text{1 cm}}\right) \, \left(\dfrac{R_0}{ \text{50 AU}}\right)^{2}.
\end{equation}

\subsection{Numerical methods}

The numerical methods are identical to those used by \citet{riols17b}. Simulations are performed with the Godunov-based PLUTO code, adapted to highly compressible flow \citep{mignone2007}, in the shearing box framework. The box has a finite domain of  size $(L_x,L_y,L_z)$, discretized on a mesh of $(N_X,N_Y,N_Z)$ grid points. The numerical scheme uses a conservative finite-volume method that solves the approximate Riemann problem at each inter-cell
boundary. It conserves quantities like mass, momentum,
and total energy across discontinuities. The Riemann problem is handled by the HLLC solver, suitable for compressible flows. An orbital advection algorithm is used to increase the computational speed and reduce numerical dissipation. Note that the
heat equation Eq.(\ref{heat_eq}) is not solved directly, since the
code conserves total energy.  Our unit of time is $\Omega^{-1}=1$, our unit of length is $H=1$, while the surface density is fixed equal to $\Sigma=1.88$. 

For details of how we calculate the 3D self-gravitating potential see \citet{riols17b} and \citet{riols17c}. The method was
tested on the computations of 1D stratified disc equilibria, as well as their linear stability, to ensure that the
implementation is correct (see appendices in \citet{riols17b}).  The boundary conditions are periodic in $y$ and shear-periodic in $x$. In the vertical direction,  we use a standard outflow condition for the velocity field and assume an hydrostatic
balance in the ghost cells for pressure, taking into account the large
scale vertical component of  self-gravity (averaged in $x$ and $y$).  Finally, the boundary conditions for the
self-gravity potential, 
in Fourier space, are:
 \begin{equation}
\dfrac{d}{dz} \Phi_{k_x,k_y} (\pm {L_z}/{2}) = \mp k\Phi_{k_x,k_y} (\pm {L_z}/{2}).
\end{equation} 
where $\Phi_{k_x,k_y}$ is the horizontal Fourier component of the potential and $k_x$, $k_y$ are the radial, azimuthal wavenumbers and $k=\sqrt{k_x^2+k_y^2}$. This condition is an approximation of the Poisson equation in the limit of low density\footnote{Indeed if the density is reduced to zero (vacuum condition), the Poisson equation is simply $\dfrac{d^2}{dz^2} \Phi_{k_x,k_y}-k^2 \Phi_{k_x,k_y}=0$. which has solutions $\propto e^{-kz}$ when $z\rightarrow +\infty$ and $\propto e^{kz}$ when $z\rightarrow -\infty$  }.
In addition, we enforce  a density floor of $10^{-4}\, \Sigma/H$ which
prevents the timesteps getting too small due to evacuated regions near
the vertical boundaries. 

For the dynamics of the dust, we use the method described and tested in Appendix A of \citet{riols18c}. In brief, we employ
a HLL Riemann solver to compute the density and momentum flux at cell interfaces. The drag force is treated as a source term in the right hand side of the second order Runge-Kutta solver. The time step is adapted to take into account the dust dynamics and the drag force. We implemented a version of the FARGO algorithm for the dust components, which splits off their mean orbital advection motion.  

{Finally, the gas is replenished near the midplane so that the total mass in the box is maintained constant. The source term in the mass conservation equation is
\begin{equation}
\varsigma(z,t)=\dot{\rho}_i(t)\exp{\left(-\dfrac{z^2}{2z_i^2}\right)}, 
\label{injection}
\end{equation}
where $\dot{\rho}_i(t)$  is the mass injection rate  and $z_i=H$ is a parameter that corresponds to the altitude below which most mass is replenished.}
  We checked that the mass injected at each orbital period is negligible compared to the total mass (less than 1\% per orbit). If not explicitly mentioned, we enact a similar replenishment for the dust. We checked also that this addition of mass does not change the main results of the paper.  

\subsection{Simulation setup and parameters}
\label{sim_setup}

The large-scale waves excited by GI have 
radial lengthscales $\lambda \gtrsim H\,Q$. In order to capture these waves, while affording reasonable resolution, we use a box of intermediate size $L_x=L_y=20 \, H$ where $H=c_{s_0}/\Omega$. The vertical domain of the box
spans $-3\,H$ and $3\,H$.  We use various resolutions, from  3  to 26 points per $H$ in the horizontal directions. 
For all simulations presented in this paper, the heat
capacity ratio is fixed at $\gamma=5/3$ and the cooling time at $\tau_c=20 \Omega^{-1}$. 

When running GI simulations with gas only,  we start from a
polytropic vertical density equilibrium, computed with an initial Toomre $Q$ slightly larger
than 1. The calculation of this equilibrium is detailed in the
appendix  of \citet{riols17b}. Non-axisymmetric density and velocity
perturbations of finite amplitude are injected to trigger the
turbulent state. For the dust runs, the initialization
is detailed in Section \ref{init_dust}: we use Stokes number between  0.0016 and 0.16 and initial dust-to-gas ratio of 0.0035 for each dust species. 

\subsection{Diagnostics}
\label{diagnostics}
To analyse the statistical behaviour of the turbulent flow, we define the standard  box average
\begin{equation}
\left<X \right>=\frac{1}{V}\int_V X\,\, dV, 
\end{equation}
where $V=L_x L_y L_z$ is the volume of the box. We also define the horizontally averaged vertical profile of a dependent variable: 
\begin{equation}
\overline{X}(z)=\dfrac{1}{L_xL_y} \int\int X\,\, dxdy.
\end{equation}
We also introduce the cross-correlation  $\star$ of two functions (integrated or averaged over $z$)
\begin{equation}
f\star g=\dfrac{1}{L_xL_y} \int \int f\left(x'+x,y'\right)\,g(x',y') dx'dy'
\end{equation}

\section{Simulation results}

\label{sec_withdust}

\subsection{Hydrodynamical gravitoturbulence}

Before we include the dust components,
 we first compute pure gaseous gravito-turbulent states similar to
 those of  \cite{riols17b} in the shearing box for different
 resolutions (from 3 to 26 points per $H$ in the horizontal
 directions). These serve as our initial conditions for the multifluid
 runs displayed in the following subsections. 

We start by analysing  some properties of these
 states. The strength and saturation of the gravito-turbulence is
 fixed by the cooling time $\tau_c$, which corresponds to the key
 control parameter.  In such turbulent  
flows the time-averaged stress to pressure ratio follows the \citet{Gammie2001} relation:  
\begin{equation}
\label{gammie_eq}
\alpha\simeq\dfrac{1}{q\Omega (\gamma-1)\tau_c} = \dfrac{1}{\Omega \tau_c}  \quad (\gamma=5/3, \, q=3/2) 
\end{equation}
In this paper we focus on the case  $\tau_c=20\, \Omega^{-1}$. The reader may refer to  \cite{riols17b} and \cite{riols17c} (Section 3.1) to obtain a detailed analysis of related simulations and more information about the turbulent properties. 
For $\tau_c= 20\,\Omega^{-1}$, the turbulence is supersonic, highly compressible and characterized by large-scale spiral density waves, particularly strong in this cooling time regime. On top of these structures, small-scale motions driven by a parametric
instability involving inertial waves attack the spiral wave fronts \citep[see][]{riols17b}.  Note that the resolution required to capture
this instability  is about $10$ points per $H$. However, we emphasize
that even for a resolution of 26 points per $H$, the smallest scales
of the parasitic inertial modes are probably not resolved, given that
it favours the smallest of scales. 

\begin{figure}
\centering
\includegraphics[width=\columnwidth]{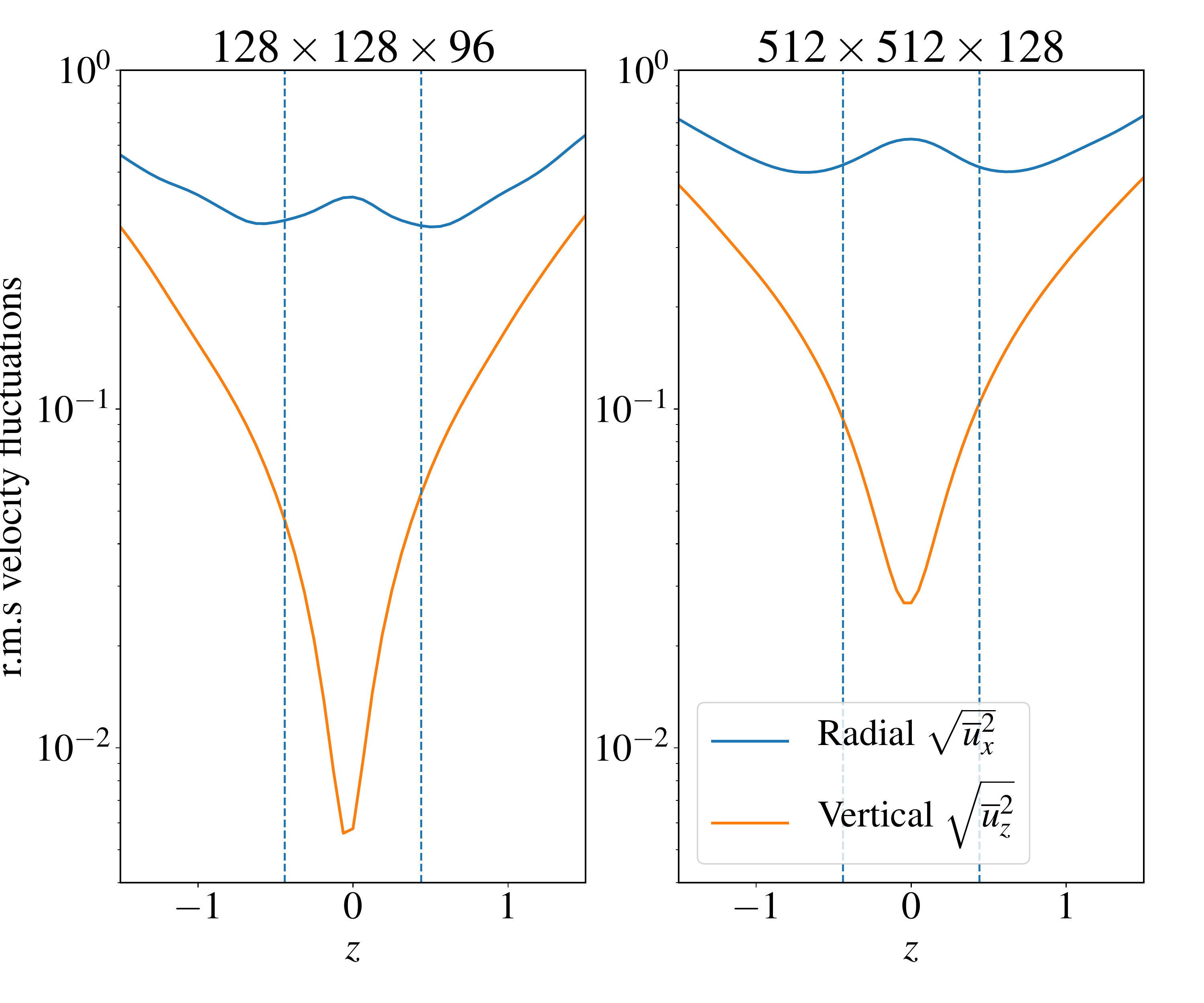}
 \caption{Mean vertical profiles of the gravito-turbulent r.m.s velocities, normalized by $c_{s_0}$.
  The quantities are averaged over time and horizontal plane, with resolution $512 \times 512 \times 128$ (left) and $128 \times 128 \times 96 $ (right).  {The time-average is done for $40\, \Omega^{-1}$ for the high resolution run and $100\, \Omega^{-1}$ for the low resolution run.} The dashed vertical lines delimit the self-gravitating disc scale-height $H_g\simeq 0.44 H$.}
\label{fig_rms}
\end{figure}

An important quantity to characterize and quantify the diffusion of solid particles in turbulent flows is the r.m.s velocity of the gas $\mathbf{u}_{\text{rms}}(z)$=$(\overline{\mathbf{u}^2})^{1/2}$.  We show in Fig.~\ref{fig_rms} the vertical profiles of the horizontally averaged r.m.s velocity  in the $x$ and $z$ directions, for runs with a resolution of 26 and 6.5 points per $H$ in the horizontal directions (respectively $N_X=N_Y=512$, $N_Z=128$ and $N_X=N_Y=128$, $N_Z=96$).  Under some approximations,  these quantities can be related to the diffusion coefficients in the radial and vertical directions and are important  to characterise the level of dust settling (see Section \ref{diffusion_model}). 
We show that the vertical and radial rms velocity increases with $z$. This profile
results from the combination of the poloidal roll motions that
accompany spiral waves at $z\lesssim H$ (see \citet{riols18}),
and small-scale inertial modes attacking these spirals at all
altitudes, but with some predominance at $z \gtrsim H$.  For a given
altitude, the radial and vertical rms velocities are stronger at the
higher resolution. We interpret this difference as a consequence of
the small-scale inertial waves, triggered at high resolution, but marginally excited at  resolution  $N_X=N_Y=128$. 

\subsection{Initialization of the dust and settling time}
\label{init_dust}
In order to simulate the dust motions unproblematically, 
we start with the gravito-turbulent state presented above, and introduce grains with initial distribution at $t=0$
\begin{equation}
\rho_d(t=0)=\rho_0\exp\left(-\dfrac{z^2}{2 H_{d_0}^2}\right), 
\end{equation}
with $H_{d_0}=0.5 \,H$ and $\rho_0$ a constant evaluated so that the ratio of surface densities $\Sigma_d/\Sigma$ is $0.0035$ for a single species (or size).  The dust velocity is initially unperturbed Keplerian motion.  We first conducted simulations at low and intermediate resolution  ($N_X=N_Y \leq 256$), for which we integrate simultaneously the motion of \emph{five} different grain sizes with Stokes numbers in the midplane 0.16, 0.06, 0.016, 0.006 and  0.0016.  We then computed two distinct high resolution ($N_X=N_Y = 512$) simulations, initialized from the same gravito-turbulent state, the first one containing particles with Stokes numbers 0.016 and 0.006,  the other containing particles with $\text{St}=0.16$ and $0.06$. 
{Also, for simplicity, the dust mass distribution is initially independent of the particle size, which is not the case in real protoplanetary discs.  However we checked that the  dust back reaction onto the gas has no important impact on the simulation outcome (see Appendix \ref{appendixB}). The initial mass distribution is then irrelevant for the dust dynamics in our problem and one can re-normalize the dust density by any given value.}   \\

Note that for a given size, the dust-to-gas ratio is not necessarily
realistic, though the total dust surface density is $0.0175$ the gas
surface density, which is not unreasonable. . 

\begin{figure}
\centering
\includegraphics[width=\columnwidth]{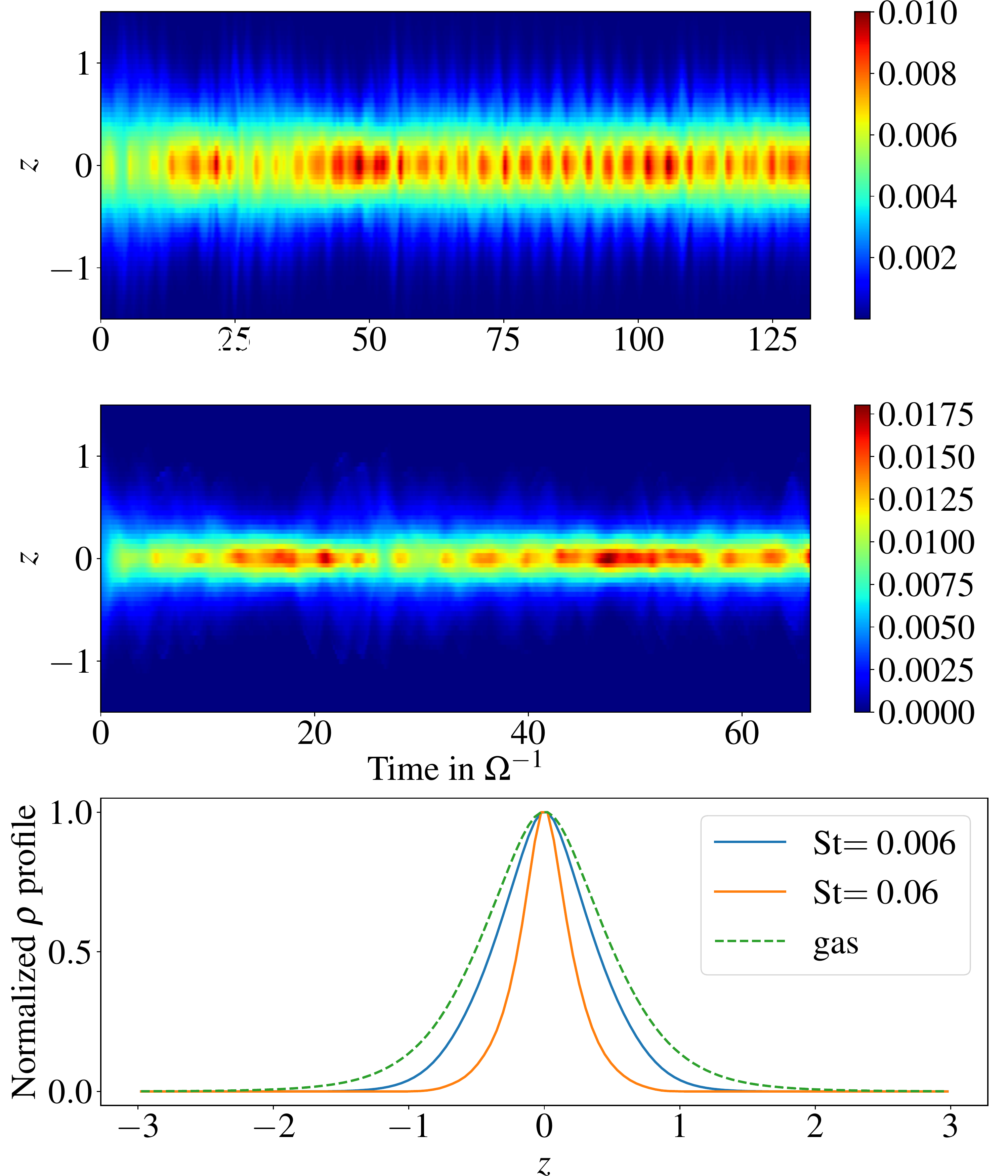}
 \caption{Top and centre panels:  space-time diagram ($z$,$t$) showing the dust density distribution,  averaged in $x$ and $y$, for $\text{St}=0.006$ and  $\text{St}=0.06$ respectively. Bottom panel: dust density profiles averaged in time and normalized to the midplane density for $\text{St}=0.006$ and  $\text{St}=0.06$. As a comparison, the green dashed curve describes the gas density profile.}
\label{fig_zprofile}
\end{figure}

Once the dust is initialized, its time and horizontally-averaged density profiles converge  toward
a steady state after a characteristic period of time dependant on the Stokes
number.  Fig.~\ref{fig_zprofile} (top and center panels) shows the
time evolution of the averaged dust density profile (in $x$ and $y$)
for $\text{St}=0.006$ and $ \text{St}=0.06$, computed from the high
resolution runs. Initially, large grains ($\text{St}=0.06$)  fall towards
the mid-plane very rapidly, within a time proportional to $
\Omega^{-1}/\text{St}\sim 15 \, \Omega^{-1}$ 
\citep{dullemond04}.  Afterwards, turbulent diffusion and mixing
emerge and ultimately balance the gravitational settling. The
mean vertical profile of the smaller grains ($\text{St}=0.006$)  does not
seem to evolve significantly during the simulation because the dust layer
is already close to equilibrium initially. However, as the space-time
diagram makes clear, on short times the vertical profiles are quite
dynamic and, in the case of small dust especially, consist of
{quasi-periodic} vertical compressions and rarefactions, clearly associated
with the spiral wave dynamics.

Note that the high resolution simulations are run for a relatively short
time ($\lesssim 100\, \Omega^{-1}$) due to the large computing
resources they demand. Nevertheless, this time remains longer or
comparable to the settling time for most of the Stokes numbers probed.
Lower resolution simulations are run for $\sim 150 \, \Omega^{-1}$  and we checked
that no significant variation of the dust dynamics occurs during  this time.

\subsection{Dust settling and vertical dynamics}
%begin{figure}
%\centering
%\includegraphics[width=\columnwidth]{FIG_GIdust_rms.pdf}
% \caption{}
%\label{fig_contrast}
%\end{figure}
\begin{figure}
\centering
\includegraphics[width=\columnwidth]{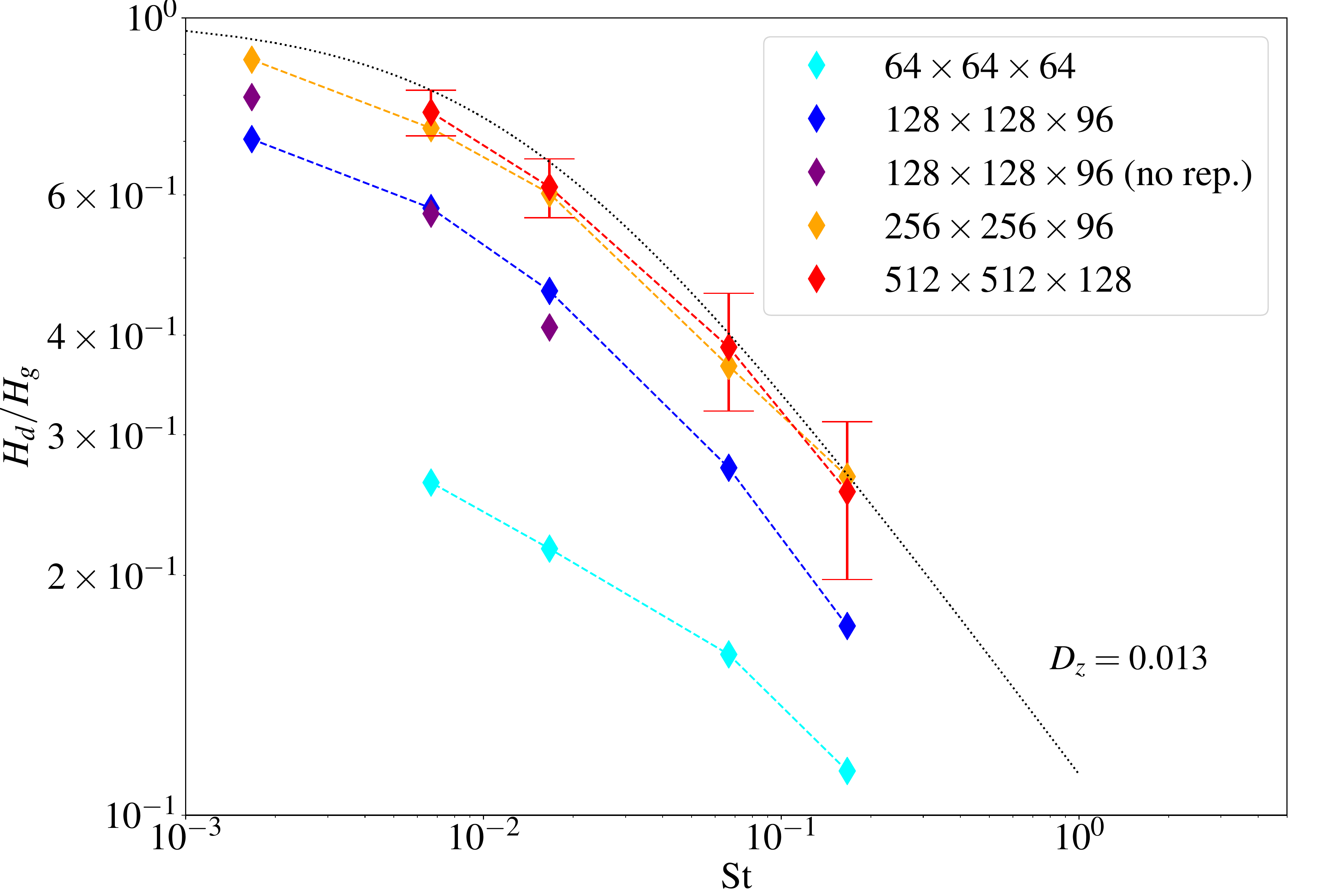}
 \caption{Mean ratio of dust  to gas heightscale $H_d/H_g$ as a function of the Stokes number for different resolution, measured from numerical simulations, with the definition given by Eq.~\ref{eq_defHd}. {In the high resolution runs we add error bars whose width corresponds to the standard deviation 
from the time-averaged $H_d/H_g$.} The purple diamonds account for a simulation without replenishment of the dust. The black dotted line is the theoretical prediction from the simple advection-diffusion model described in Section \ref{diffusion_model} using a diffusion coefficient $D_z=0.013$.}
\label{fig_Hd}
\end{figure}

\subsubsection{Vertical  density profiles and scaleheights}
\label{density_profiles}
We characterize the long-term dust vertical equilibrium and estimate
its typical scale height as a function of the Stokes
number. Fig.~\ref{fig_zprofile} (bottom panel) shows the mean vertical
density profiles, averaged in time ({during $130 \,\Omega^{-1}$ for  $\text{St}=0.006$ and  $60\,\Omega^{-1}$ for $\text{St}=0.06$}) and obtained in the high resolution runs $(512 \times
512 \times 128)$. For comparison we superimpose the gas vertical
density profile (dashed green line); though not strictly a gaussian, this curve can
be fitted rather well with one, with width $H_g \simeq 0.44 H$. 

The dust density profiles can also be approximated by
gaussians but with a width smaller than $H_g$.  
 We define the dust scaleheight $H_d\,(\text{St})$ as the altitude $z$ such that
\begin{equation}
\label{eq_defHd}
\rho_d (z=H_d) = \rho_d (z=0) \, e^{-\frac{1}{2}} \simeq 0.6\, \rho_{d0}.
\end{equation}
For each species, we measure this scaleheight and display the {time-averaged} dust to
gas ratio $H_d/H_g$ in Fig.~\ref{fig_Hd} for various resolutions. 

First we see that, independently of resolution, the size of the dust layer increases with decreasing Stokes number. This is to be expected, 
because small dust particles are less sensitive to gravitational settling and will tend to follow the turbulent gas motion.  
At larger St, the ratio $H_d/H_g$  depends on St$^{-1/2}$, a result
that has been obtained in other simulations coupling dust and
turbulent gas {\citep{fromang06b,okozumi11,zhu15,yang18,riols18c}}. This
dependence  can be understood, in a rather crude way,  within the
framework of a simple diffusion theory \citep[][see Section
\ref{diffusion_model}]{morfill85,dubrulle95}, where the vertical
equilibrium is set by the balance between the gravitational settling
and turbulent diffusion. 

Second,
the absolute values of $H_d/H_g$ increases with the grid
resolution. The reason of this dependence may be attributed to the
difficulty in simulating the parametric instability, which excites small-scale modes
that may enhance diffusion of
dust particles. Lower resolution runs do not adequately capture these small-scale modes and hence
the diffusion they bring to bear on the dust. Note, however, that convergence does seem to be achieved
for a resolution greater than 13 points
per $H$ (the case with 13 or 26 points in the horizontal directions showing no major difference). 

Third, the size of the dust layer is large for mm to
cm particles ($\text{St}\simeq 0 .0016 $ and $0.016$),
larger than $0.85 H_g$ and $0.6 H_g$, respectively.
This is very similar to what magnetorotational turbulence with a zero-net 
vertical  field can achieve \citep{fromang06b}.  These layer thicknesses are
interesting since they can be directly measured in cases where the disc is observed
edge-on. Indeed, the spatial resolution of
instruments like ALMA  is sufficient to resolve vertical scales less than
$H$ at distances of a few tens of AU (see discussion in Section 4). 

{Finally, as mentioned already in Section \ref{init_dust}, Fig~.\ref{fig_zprofile} indicates that the dust midplane density varies quasi-periodically (with period of a few $\Omega^{-1}$). Concurrently, 
the dust layer undergoes vertical compression and expansion, which are clearly correlated with the variations of the gas midplane density. Inevitably these oscillations lead to variations in the dust scale height. We thus quantify, for the high resolution runs, the typical deviations of $H_d$ (denoted $\delta H_d$) and  the ratio $H_d/H_g$ (denoted $\delta H_d^*$) from their temporal averages. We find that $\delta H_d \simeq 0.19, 0.16, 0.15$ and $0.13\, H_d$ respectively for $\text{St}=0.16,0.06,0.016 $ and 0.006. In the same order, we find $\delta H_d^*=0.17,0.12,0.1$ and $0.07\,(H_d/H_g)$. The last values are used to calculate the error bars in Fig.~\ref{fig_Hd}. Thus the deviations (or oscillations) remain relatively 
small compared to the mean values and will be probably undetectable by current instruments measuring the dust scaleheight.}

\subsubsection{Settling model and diffusion coefficients}
\label{diffusion_model}

We next apply a simple diffusion model \citep{dubrulle95} to explain the equilibrium dust
scaleheights measured in the previous section. The model has its limitations;  in particular, it assumes that
turbulent eddies sizes are less than $H$, whereas the
GI vertical rolls occur on scale similar or larger than $H$.
Nevertheless, assuming that the theory is marginally applicable, 
we find (see Appendix \ref{appendixA}) that the dust to gas scaleheight  ratio is  
\begin{equation}
\label{eq_Hdgauss2}
\dfrac{H_d}{H_g} =\left(1+\dfrac{\text{St}\,\Omega f_c (s+1)  \,H_g^2}{D_z }\right)^{-1/2}.
\end{equation}
with $f_c\approx1.3$ a coefficient related to the compressibility of the flow, $s\simeq2.77$ a coefficient related to the settling due to  self-gravity and $D_z \simeq  \langle \overline{v_z^2} \rangle $ $\tau_{\text{corr}}$  a constant and uniform diffusion coefficient encapsulating turbulent transport. 

\begin{figure}
\centering
\includegraphics[width=\columnwidth]{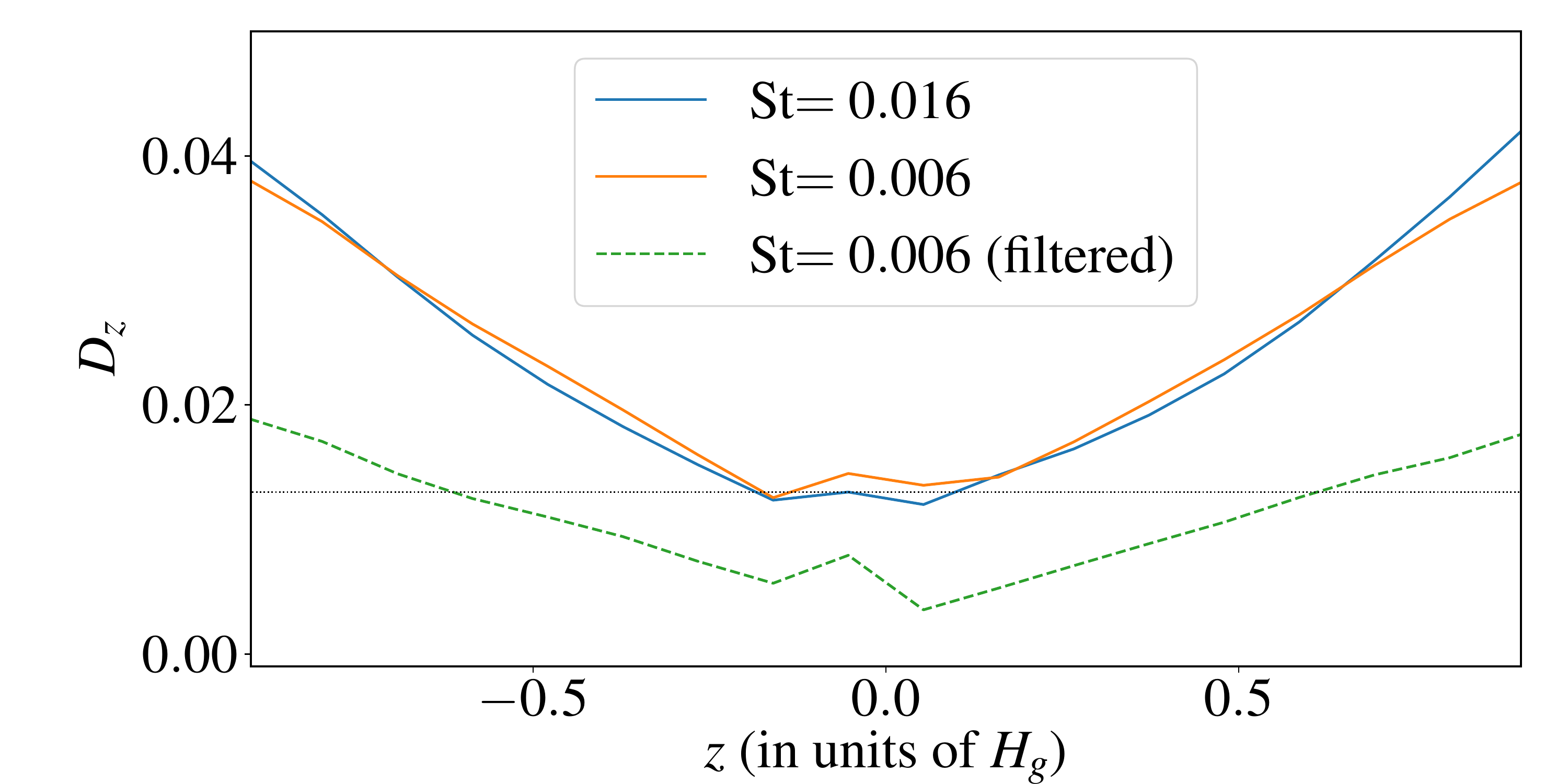}
 \caption{Vertical diffusion coefficients measured in the high resolutions simulations with $N_X=N_Y=512$. The blue and orange curve are respectively computed  for $\text{St}=0.016$ and $\text{St}=0.006$. The green curve is the diffusion coefficient computed by filtering out all the modes $k_y>2 \pi /L_y$ in the calculation of the averaged product $\langle\overline{\delta \rho_d \,\delta v_{z_d}}\rangle$. This gives an estimate of the diffusion produced by large scale spiral waves only. Note that the small asymmetry about the midplane is potentially due to the fact that the averaging procedure  is done over a rather short time  ($\sim 100\, \Omega^{-1}$). }
\label{fig_Dz}
\end{figure}

We could, of course, apply Eq.~\eqref{eq_Hdgauss2} to Fig.~\ref{fig_Hd} and 
find the $D_z$ predicted by the theory in each case. Instead, we calculate $D_z$ 
directly from the simulations and subsequently check how well the diffusion theory
does in reproducing Fig.~\ref{fig_Hd}. We compute the vertical diffusion coefficient from the high
resolution simulation data, by averaging in time over
 $100\,  \Omega^{-1}$, the quantity
\begin{equation}
\label{def_Dz}
D_z = -\dfrac{\langle\overline{\delta \rho_d \,\delta v_{z_d}}\rangle}{\left \langle \overline{\rho} \dfrac{\partial}{\partial z} \left(\dfrac{\overline{\rho_d}}{\overline{\rho}}\right)\right \rangle },
\end{equation}
(see Appendix \ref{appendixA}). Figure \ref{fig_Dz} shows $D_z$ calculated this way
for two different Stokes numbers $\text{St}=0.016$ and $\text{St}=0.006$, as
a function of $z$. In the midplane we find $D_z \simeq 0.013$ roughly for both
{Stokes numbers}, but note that $D_z$ increases with altitude $z$, following
the r.m.s vertical velocity in Fig.~\ref{fig_rms}. Despite this increase,
the hypothesis of constant $D_z$ does hold for $z\ll H_g$.  If next we
insert this constant numerical value in place of the diffusion coefficient in
Eq.~\ref{eq_Hdgauss2}, we obtain a ratio $H_d/H_g$ that reproduces
that measured in  the high resolution simulation (see
dotted black line in Fig.~\ref{fig_Hd}). We hence conclude that, 
to a first approximation, the several turbulent gas flow features
acting on the dust work together diffusively, at least on long times.
(On short scales, of course, the situation is more interesting and
dynamic, as the top two panels of Fig.~\ref{fig_zprofile} indicate.)

\subsubsection{The relative contributions of vertical rolls and small-scale
turbulence to diffusion}

\begin{figure}
\centering
\includegraphics[width=1.02\columnwidth]{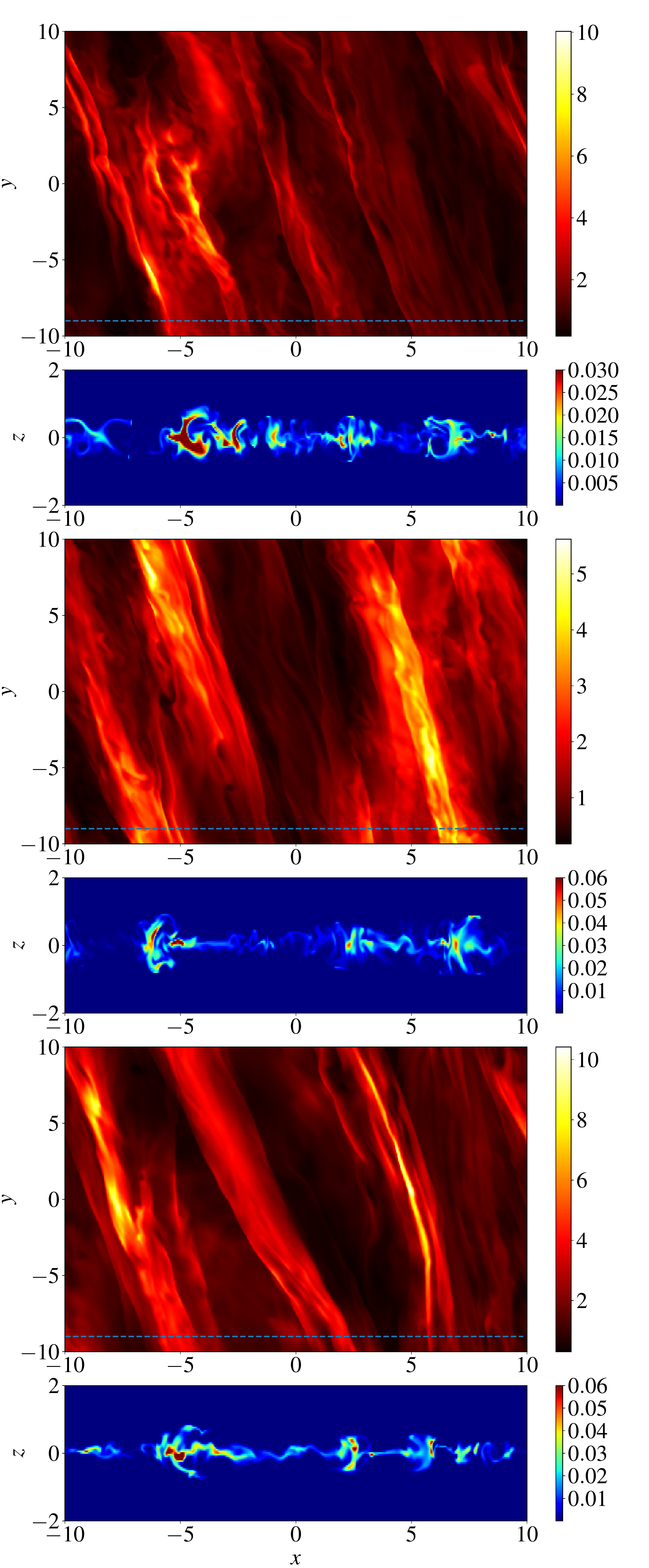}
\caption{Snapshots of the gas density in the horizontal plane and the dust density in the poloidal plane  for $\text{St}=0.06$ and for three
 different times {(from top to bottom : $27\, \Omega^{-1}$, $31 \Omega^{-1}$ and $51 \, \Omega^{-1}$)}. The dashed lines in the former denote the location of the poloidal planes.}
\label{fig_rolls}
\end{figure}

In the previous subsection we demonstrated that the
gravitoturbulence can effectively halt the settling of dust grains.
Now we determine what features of the flow are responsible
for the vertical diffusion of the particles. In particular, which is
more important: small-scale inertial wave turbulence (difficult to
simulate because of steep resolution requirements), or large-scale
vertical rolls (somewhat more easy to simulate, especially in global
set-ups).  

We begin by concentrating on the large-scale vertical circulation.
As shown by \citet{riols18}, in stratified
atmospheres with a mean entropy gradient, these motions are quite
generally triggered
by baroclinic effects and are composed of a pair of counter-rotating
rolls of size $\simeq H_g$, travelling in the horizontal direction
with the wave. In severe spiral shocks, vertical flows arise also from
hydraulic jumps (Boley and Durisen 2006).

In Fig.~\ref{fig_rolls}, we show the gas distribution in
the horizontal plane and the corresponding dust distribution in a
poloidal plane $(y=-9H)$ for $\text{St}=0.06$ at 3 different times. At
the location of each spiral wave, the dust distribution forms vertical
arcs that locally reach $z \simeq H$. These arcs are clearly the
result of dust lifted up by the large-scale rolls. Note that these arcs are not 
necessarily symmetric with respect to the wave front and are stretched in a privileged  direction. 
Clearly we see a dynamic transport of dust vertically, but it is not guaranteed that, 
cumulatively, these arcs lead to an appreciable average vertical diffusion of dust
(and thus impact on the ratio
  $H_d/H_g$).  Indeed the inter-arms spirals
regions are highly sedimented at the same time these arcs are active.  

To make further progress and to develop a more quantitative approach, 
{we remove the contribution of axisymmetric modes and 
non-axisymmetric modes with $k_y> 2 \pi/L_y$}  in the calculation of
the vertical diffusion coefficient
$D_z$. 
This is equivalent to filtering out the small-scale inertial wave turbulence
and keeping only the large-scale spiral vertical rolls (with $k_y=2 \pi/L_y$)
 in the product $\langle\overline{\delta \rho_d
  \,\delta v_{z_d}}\rangle$. The related diffusion coefficient $D_{z_{\text{filt}}}$ 
is shown in dashed green in Fig.~\ref{fig_Dz}. We see that the filtered
quantity contributes 30 \% of the diffusion coefficient in the
midplane  ($D_{z_{\text{filt}}} \simeq 0.0035$) and rises to 50 \% in
the corona. Therefore, both the spiral vertical motions and the
small-scale inertial waves contribute
to the dust diffusion. 

In fact, this result may have been expected from
Fig.~\ref{fig_Hd}, which shows the dependence of $H_d/H_g$ on resolution. For
the lowest resolution $(N_X=N_Y=64)$, the small-scale turbulence is
not properly resolved and its impact in the dynamics diminished, as a consequence; 
thus the vertical diffusion is accomplished primarily by the
vertical rolls, and indeed we see immediately that the dust scale height drops
significantly and is well-approximated by a theory using the filtered
diffusion coefficient $D_{z_{\text{filt}}}$. At better resolution the small-scale
turbulence is better described, vertical diffusion increases as a result, and the
dust thickness increases. 

\begin{figure}
\centering
\includegraphics[width=\columnwidth]{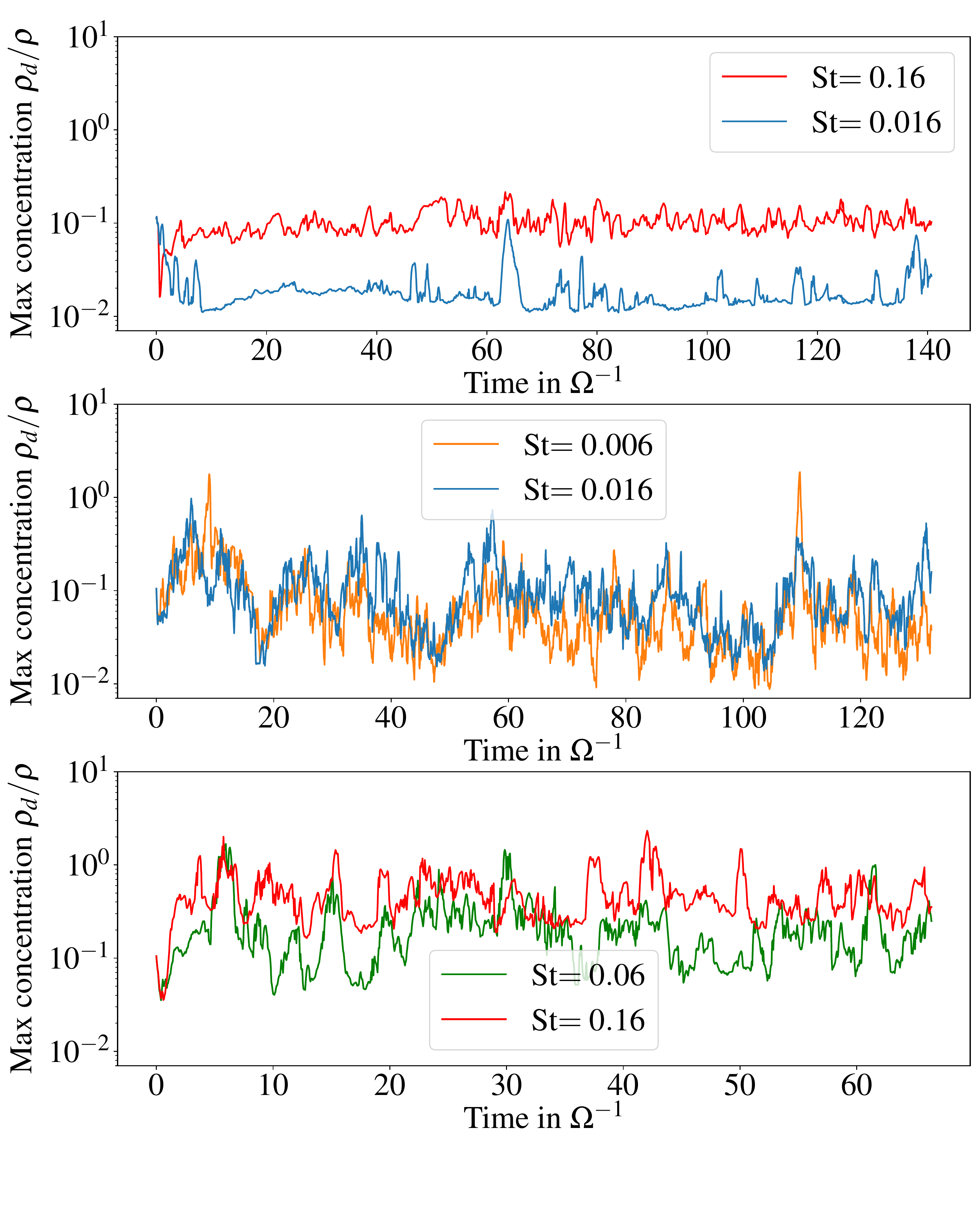}
 \caption{Maximum dust concentration in the box for different Stokes numbers. These are computed for resolutions $128 \times 128 \times 96$ (top panels) and  $512 \times 512 \times 128$ (middle and lower panels)}
\label{fig_co2}
\end{figure}

\begin{figure}
\centering
\includegraphics[width=\columnwidth]{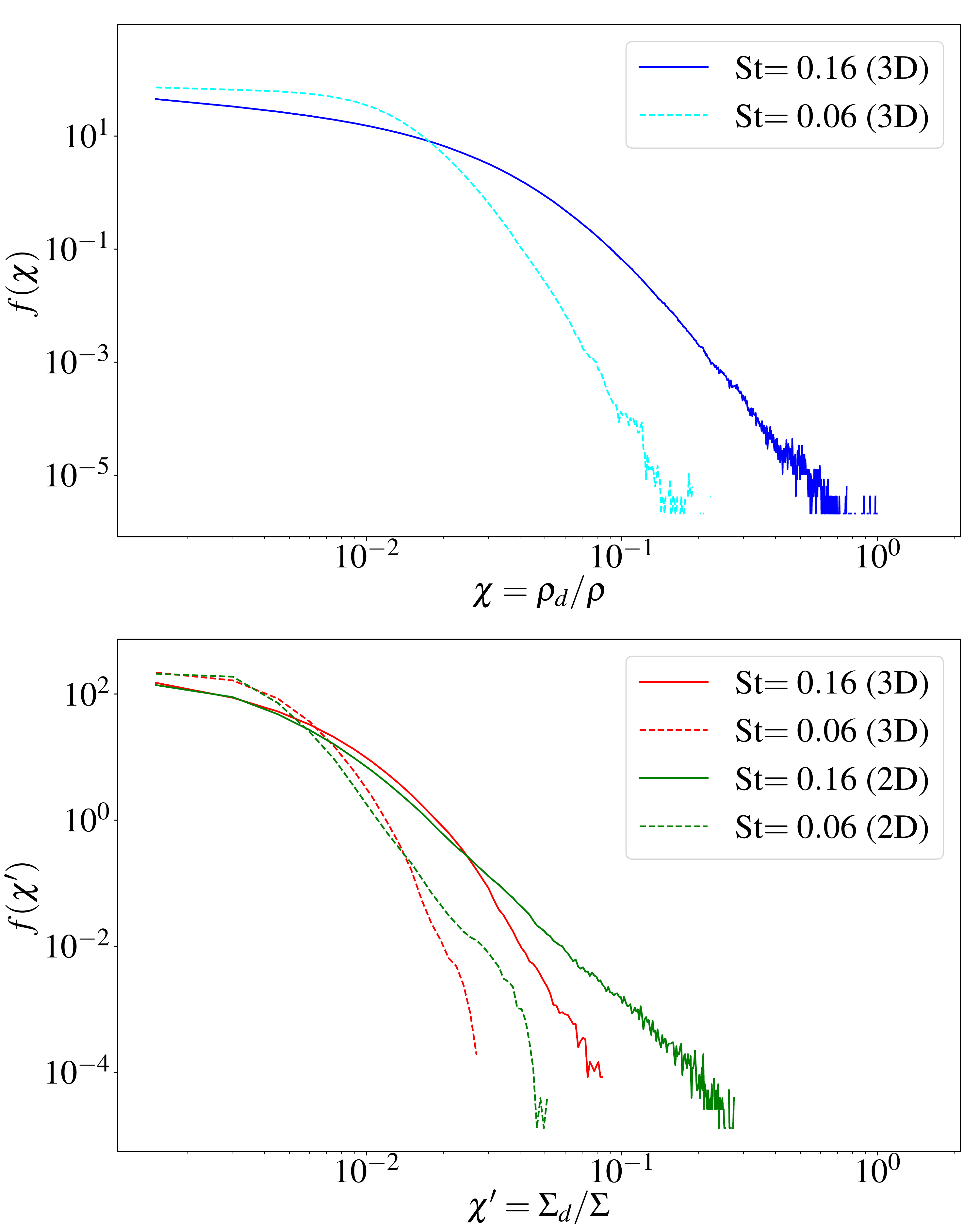}
 \caption{Probability distribution functions $f$ for concentration events. The top panel shows $f(\chi)$ computed from 3D simulations,
 where  $\chi=\rho_d/\rho$.  The bottom panel shows $f(\chi')$ associated with 2D simulations (in green) and 3D simulations (in red) 
where $\chi'=\Sigma_d/\Sigma$. The dotted lines are for $\text{St}=0.06$ and the solid lines are for $\text{St}=0.16$. }
\label{fig_pdf}
\end{figure}

\begin{figure*}
\centering
\includegraphics[width=\textwidth]{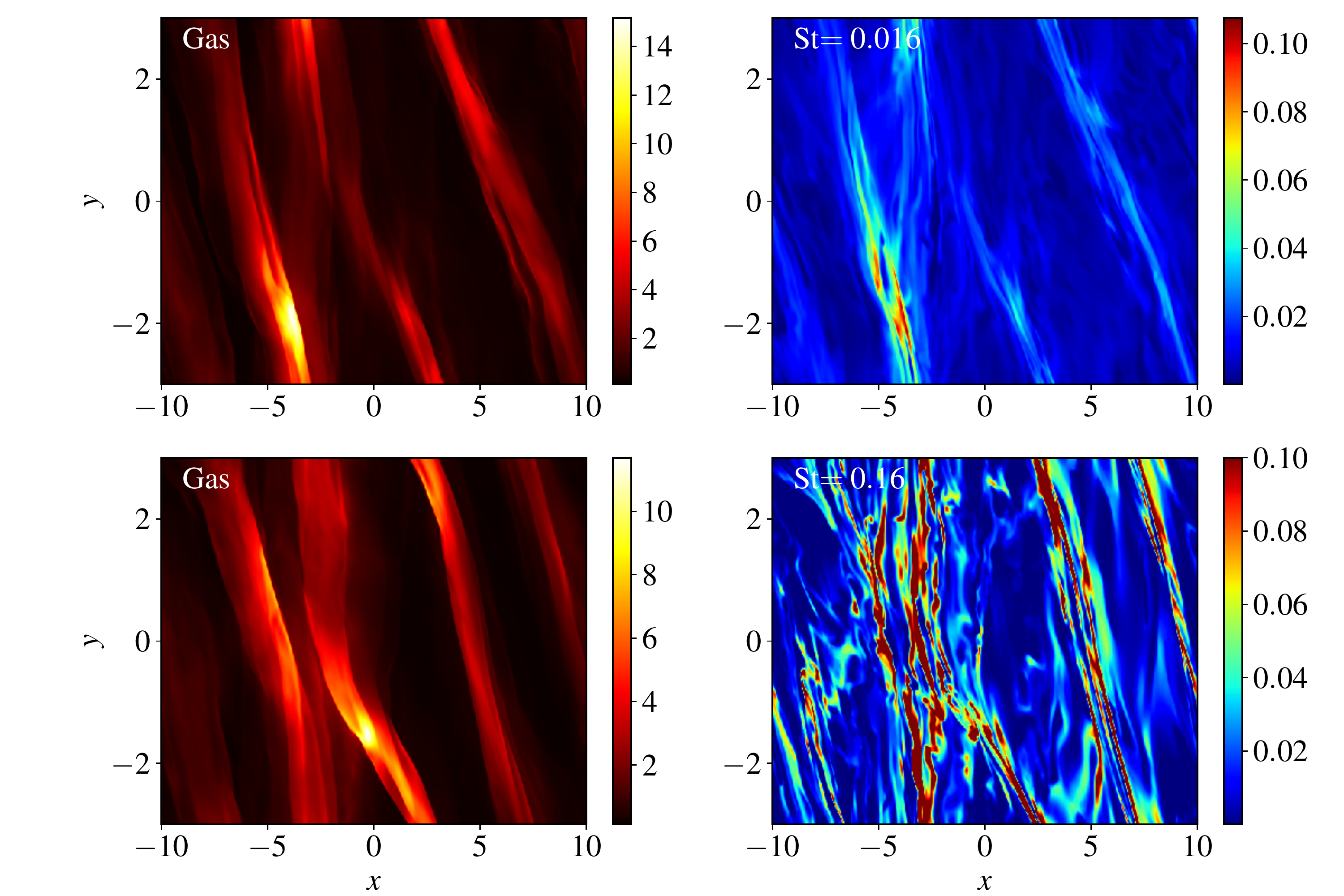}
 \caption{Left panels: snapshots of gas pressure in the midplane at two different times. The top panels are taken at $t=48\Omega^{-1}$ while bottom panels are taken at $t=50 \Omega^{-1}$ when the dust to gas ratio reaches a local maximum (we remind the reader that top and bottom are different simulations with different grain size, but with the same initial conditions for the gas) . Right panels correspond to the dust density in the midplane for $\text{St}=0.016$ (top) and $\text{St}=0.16$ (bottom).}
\label{fig_co}
\end{figure*}

\begin{figure}
\centering
\includegraphics[width=\columnwidth]{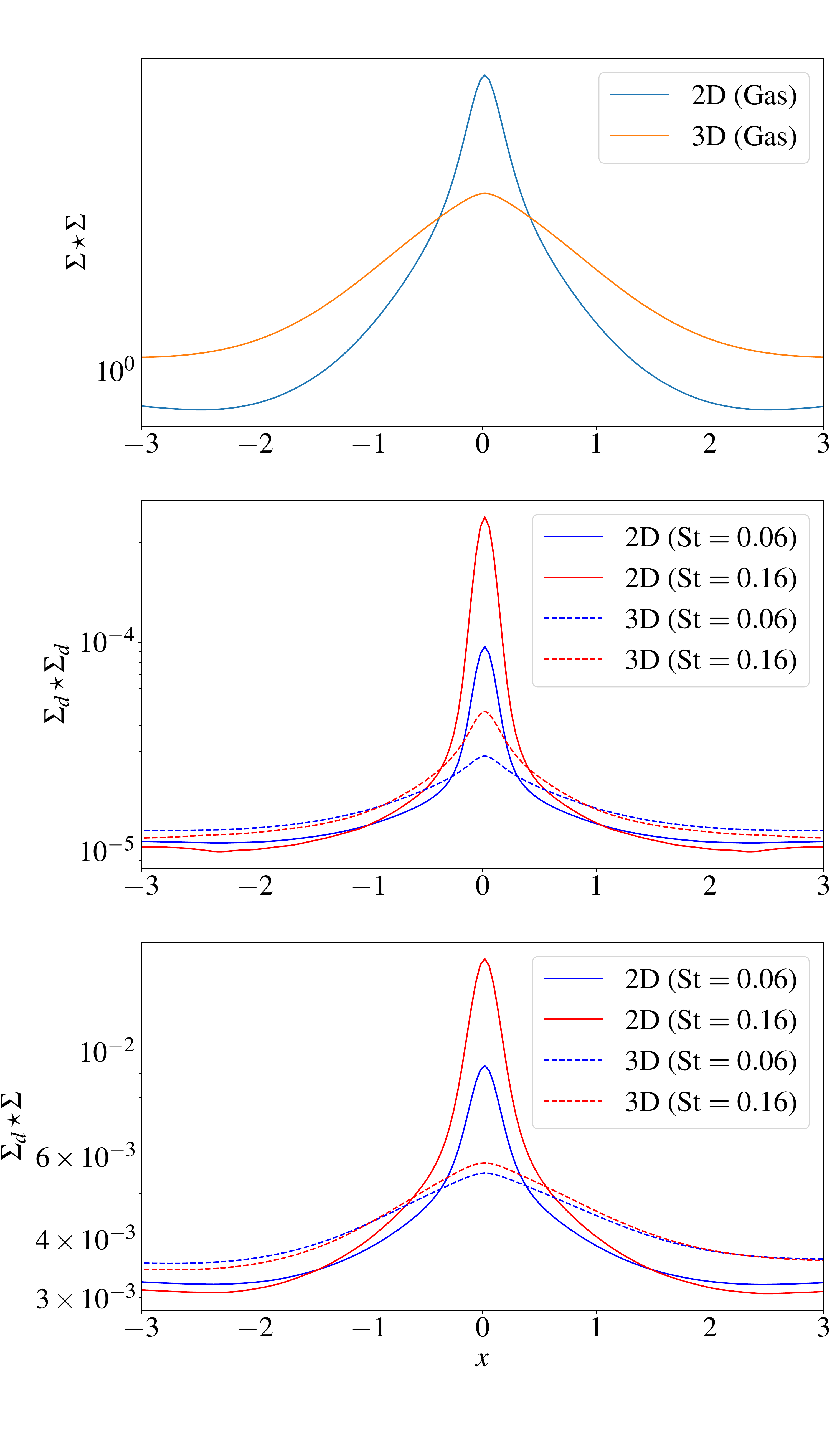}
 \caption{Top and centre panels are, respectively, the autocorrelation functions of the gas and the dust surface density in the radial direction. 
 Bottom panel shows the cross-correlation between gas and dust surface density in the radial direction. Each correlation function is averaged in time. }
\label{fig_corr}
\end{figure}

\subsection{Dynamics in the horizontal plane}

\subsubsection{Concentration events}
\label{concentration_events}
In this section we analyse the statistics of concentration events,
especially those that lead to high dust to gas ratios
$\rho_d/\rho\gtrsim 1$.
In dust rich regions the system might trigger the
streaming instability or even the gravitational collapse of
the dust, which may be crucial stages in the planet formation process {(See Section \ref{sec_intro}).}

Before we present this subsection’s results we must emphasise that
they are \emph{resolution dependent}: specifically, the better the resolution
the more likely the dust is to be concentrated. This dependence probably
issues from two causes: (a) some of the properties of 
our simulated small-scale inertial wave turbulence are not
converged with respect to resolution, because the parametric instability can inject energy 
into extremely short scales, shorter than our grid lengths, and (b) the violation of
the pressureless fluid approximation for the dust in high resolution runs, because the stopping
time may become longer than the turbulent turn-over time on the grid. Certainly, the latter
effect will artificially enhance concentration events, and thus our high resolution 
results may best be understood
as providing \emph{upper bounds} on concentration. Perhaps more robust are the relative trends observed
and the differences between 2D and 3D simulations.

 First, we show in Fig.~\ref{fig_co2} the time-evolution
of the maximum concentration $\rho_d/\rho$ in the box. This
concentration, on average, increases with Stokes number, which is
expected from physical arguments. Small grains mostly follow the gas motion, 
whereas particles with Stokes
number $0.16$ can drift more easily toward pressure
maxima. Fig.~\ref{fig_co2} shows that in the high resolution runs, and
for $\text{St}=0.16$, the concentration of dust rarely exceeds 1 during
the first tens of orbits. Obviously such events are even less frequent
for small particles but can still occur  (for example at $t=10$ and
$t=110\, \Omega^{-1}$ for $\text{St}=0.006$). However all these events
are short and never last more than an orbit. 
Note that in our low resolution simulation, significant concentration events do not occur 
(see discussion above).

To further investigate the occurrence of particle concentration, we
show in Fig.~\ref{fig_pdf} the probability distribution function
$f(\chi)$ for concentration events, computed for two different Stokes
numbers in the midplane region ($\vert z \vert \leq 0.4 H$). If we set
$\chi=\rho_d/\rho$, the function $f(\chi)$ is obtained by counting
the number of cells within the midplane that contain a given
concentration $\chi$, at any given time. The function is averaged in time and then
normalized so that its integral over the domain of $\chi$ considered
is 1. We find again that the largest Stokes number $\text{St}=0.16$,
which corresponds to a grain of decimetre size, favours higher dust
concentration. The function has a small tail at $\chi \lesssim 1$, 
but the probability of $\chi = \rho_d/\rho \gtrsim 1$ is 
almost zero and concentration events are very rare.  

In the lower panel of Fig.~\ref{fig_pdf}, we compare this result with 2D planar simulations possessing the same Stokes numbers, cooling time, and initial surface densities. Note that the 2D simulations are performed
without a smoothed potential in the vertical direction, and thus solve
\begin{equation}
\mathbf{\nabla}^2\Phi_s = 4\pi G\rho(x,y)\delta(z) .
\label{poisson_eq}
\end{equation}
with $\delta$ the Dirac function. To make the comparison possible,  we compute the probability distribution function for the ratio of surface densities $\chi'=\Sigma_d/\Sigma$. The result is that the tail of the distribution function in 2D simulations extends to larger $\chi'$, and
hence stronger concentration events are more likely. 
One hence concludes that the inclusion of additional 3D flows 
works against the formation of dense columns of dust, probably via a combination of vertical redistribution of dust by the 
vertical rolls and small-scale inertial wave turbulence.

\subsubsection{Grain distribution within spiral waves}

Although the concentration of sub-decimetre dust grains seems to barely reach 1, it is of interest to determine
how the dust is distributed horizontally. We find that grains are mostly concentrated into the pressure maxima associated with spiral waves,
in agreement with previous work \citep{gibbons12,gibbons15,shi16}.  
To illustrate this result, we show in Fig.~\ref{fig_co} two snapshots of the gas pressure and dust density taken from the high resolution simulations. The upper right panel corresponds to $\text{St}=0.016$ at a random time, while the lower right panel  corresponds to $\text{St}=0.16$ at $t=50\, \Omega^{-1}$ when the concentration reaches a local maximum $\rho_d/\rho \simeq 1$.   Clearly small particles are well coupled to the gas and therefore display a similar density structure. Particles possessing the longer stopping time  $\text{St}=0.16$ concentrate in thin filaments, located within the spiral waves, 
and exhibiting densities two or three order of magnitude greater than the background dust density.  

To be more quantitative, we analyse the typical length-scales of the  gaseous and dust structures in the radial direction.  For that purpose, 
we  introduce the two auto-correlation functions $\Sigma \star \Sigma(x)$ and $\Sigma_d \star \Sigma_d(x)$ (see definitions in Section \ref{diagnostics}),  averaged in time during the course of the simulation.
The typical width of these functions (which is taken as $2x$ with $x$ corresponds to the radius of half their peak amplitude) account respectively for the size of the gas spiral arms and the dust structures in the radial direction.

Figure \ref{fig_corr} (top) shows $\Sigma \star \Sigma$ for different 2D and 3D runs. For a similar box size $L_x=20 H$, the spiral arms obtained in 2D are two times thinner than those obtained in 3D. For reference, we denote by $\lambda_{2d}$ and $\lambda_{3d}\simeq 2H $ these different scales associated with the gas spiral structures.  Figure \ref{fig_corr} (center panel) shows the auto-correlation functions of the dust $\Sigma_d \star \Sigma_d$ for 2D and 3D runs and for Stokes numbers St $=0.06$ and St $=0.16$. Clearly 2D dust structures are much thinner than those in 3D. The dust filaments in 2D have lengthscale between $ 0.18$ and $0.25 \lambda_{2d} $ while in 3D the structures are much wider with typical size between  $ 0.4$ and $0.8 \lambda_{3d} $ (comparable to the gas spiral arms).  We checked also that the dust is concentrated into thinner structures when the Stokes number increases, which is expected.

Finally, to understand the distribution of dust relative to the gas, we  introduce the cross-correlation function $\Sigma_d \star \Sigma$ displayed in the bottom panel of Fig.~\ref{fig_corr}.  Clearly, for all cases, the cross-correlation has a maximum at $x=0$ suggesting that the dust is trapped, on average, into the density maxima of the gas corresponding to spiral waves. More interestingly, the correlation is higher in 2D than in 3D and the typical correlation length is smaller in the 2D case ($\simeq 0.5 \lambda_{2d}$ in 2D versus $\simeq 2 \lambda_{3d}$ in 3D). Physically this means that in 3D, diffusion of dust is enhanced  and counteracts the process of grain accumulation inside spiral waves. 

\section{Discussion and Conclusions}

\label{sec_conclusions}

In this paper, we simulated dust dynamics in gravitoturbulent
accretion discs. Our special focus was on the action of
secondary 3D flows associated with the spiral waves (vertical rolls
and inertial wave turbulence), and thus we employed high resolution
vertically stratified shearing boxes, using the code PLUTO.

First, we showed that both small-scale GI motions and
large-scale vertical rolls associated with spiral waves act to diffuse
the grains in the vertical direction. We calculated
the steady-state dust scaleheights as a function of Stokes number, and
showed that a simple diffusion model  \citep[like that
employed by][]{dubrulle95} is sufficient to explain the scaleheights
measured in simulations. Note that the Schmidt number, the ratio of
turbulent viscosity  to particle diffusion coefficient
{
\begin{equation}
\text{S}_c=\nu_t/D_z =\alpha H^2 \Omega/D_z = H^2/(q(\gamma-1) \tau_c  D_z), 
\end{equation}
(assuming Eq.~\ref{gammie_eq} for $\alpha$)} is of order 4, twice that measured
in the radial direction in 2D simulations \citep{shi16}. Overall, we find
that GI significantly impedes the settling of intermediate size grains:
quasi-steady dust scale-heights are roughly half or more the accompanying
gas scale height. This is perhaps the most interesting, and most robust, result 
in this paper.

Second, we
studied the dynamics in the horizontal direction and found that
concentration of grains into pressure maxima (i.e spiral waves)  is
less pronounced in 3D than in 2D, although small-scale filamentary
structures embedded in the spiral waves still occur.
The ratio $\rho_d/\rho$ never exceeds 1 and transient concentration events 
possess timescales that barely reach an orbit for grains below
the decimetre size. We stress, however, that these results suffer from
resolution non-convergence, issuing either from our fluid model for the dust or the
difficulty in simulating 
the small-scale turbulence. 

Lastly, we showed that the typical
horizontal lengthscale of dust spiral structure in 3D is longer than that of gas
spiral arms for $\text{S}t \lesssim 0.2$, and in particular significantly longer than in
comparable 2D runs. This suggests that
additional 3D flows    
act to diffuse the grain in the radial direction and prevent its
concentration into the pressure maxima. 
In other words, the secondary vertical rolls and small-scale inertial
turbulence help `blur' the signature of the gas's spiral waves in the
horizontal dust distribution. 

Our results have several implications for young and massive
protoplanetary discs and their observations. First, they invite us to
reassess
the conclusions of previous 2D studies on the formation of planetesimals by
GI  \citep{gibbons12, gibbons15, shi16}. Three dimensional
flows disfavour the concentration of grains, via their
retardation of vertical settling and via radial diffusion. 
Consequently, these flows indirectly inhibit
the streaming instability acting on centimetre to decimetre sizes
\citep{youdin05} and the direct gravitational collapse of such grains. 

Second, and on the other hand, the inefficient sedimentation of
sub-millimetre particles could help us infer the
existence of gravito-turbulence in the outer radii of
protoplanetary discs. At these radii, non-ideal
effects, in particular ambipolar diffusion,  is believed to quench the
magneto-rotational instability \citep{fleming00,
  sano02,wardle12,bai13b,lesur14, bai15} and prevent any form of
turbulence originating from MHD effects. Thus a low level of settling
measured in observed discs is likely to be induced by
hydrodynamic turbulence such as GI {\citep[or the VSI if sufficiently strong, see][]{stoll16,lin19}} 
In the coming years, the
radio-interferometer ALMA will be able to study a large sample of `edge-on’ discs and the sufficient resolution to measure  the
dust scaleheight in these systems. The comparison between these
scaleheight and those simulated will be directly relevant to  assess the presence of GI in these discs.  
 
Finally, the 3D flows accompanying spiral arms in GI could have a direct impact on the scattered infra-red luminosity measured from observations.
We have shown that small dust particles (with stopping times much less than $\Omega^{-1}$) are lofted efficiently 
above the spiral patterns at the disc surface, and also mixed in the upper layers by small-scale turbulence. As a result the surface emission properties of the disk will be altered.

\section*{Acknowledgements}
 This project has received funding from the European Research Council (ERC) under the European Union’s Horizon 2020 research and innovation programme (Grant agreement No. 815559 (MHDiscs)) This work was granted access to  the  HPC  resources  of  IDRIS  under  the  allocation  A0060402231 made  by GENCI  (Grand  Equipment  National  de  Calcul  Intensif).  Part  of  this  work was  performed  using  the  Froggy  platform  of  the  CIMENT  infrastructure (https://ciment.ujf-grenoble.fr).

\bibliographystyle{mnras}
\bibliography{refs} %if your bibtex file is called example.bib

\appendix 

\section{Settling model}
\label{appendixA}

We use here the simple diffusion theory of \citet{dubrulle95} to estimate the dust scaleheight in  gravito-turbulence. In this model,  
it is assumed there is some `small-scale’ turbulence, with a characteristic horizontal lengthscale, a characteristic
 vertical lengthscale $\ll H$, and a timescale of order
an orbit. In addition, there is a large-scale mean component that varies on long times, much longer than an orbit,
and exhibits variations only in $z$, of an order the disk scaleheight, $H$. 

We introduce fast variables $\mathbf{x}’$ and $t’$, which vary on the turbulent scales, and slow variables $z$ and $t$,
which vary on the long mean scales \citep[see][]{latter12}. Next
all quantities are decomposed into mean and fluctuating parts
\begin{equation}
\label{eq_decompo2}
\rho_d = \overline{\rho_d}+ \delta \rho_d; \quad   \mathbf{v} = \overline {\mathbf{v}}+ \delta \mathbf{v};  \quad \mathbf{v_d} = \overline {\mathbf{v_d}}+ \delta \mathbf{v_d},
\end{equation}
with the mean parts depending only on the slow variables and the fluctuating parts depending on both slow and fast variables. 
To formally distinguish the two components we introduce the average $\overline{f}=\int f d\mathbf{x}’ dt’$, which integrates over
sufficient turbulent length and time scales so that $\overline{f}$ only depends on the slow variables, 
where $f$ is any field and $\delta f$ is the fluctuating
component of that field.

The averaged mass conservation equation (\ref{eq_mass_dust}) can be written as:
\begin{equation}
\dfrac{\partial \overline{\rho_d}}{\partial t}+\dfrac{\partial}{\partial z} \left(\overline{\rho_d}\, \overline{ v_z} + \overline{\delta \rho_d \,\delta v_{z_d}}+ \overline{\rho_d} \, \overline{\Delta v_z}\right)=0,
\label{mass_eq_dust2}
\end{equation}
where $\mathbf{\Delta v}=\mathbf{v}_d-\mathbf{v}$ is the drift velocity between dust and gas. The first term in the $z$-derivative corresponds to the advection-stretching of dust by the mean vertical gas flow (wind), which appears to be negligible in our numerical simulations. The second term is the correlation of turbulent fluctuations which is approximated by a diffusion operator in Dubrulle's theory. The third term accounts for the mean vertical drift of dust due to gravitational settling (including the self-gravity of the disc).   Using classical  assumptions, detailed in Section 5.1.2 and Appendix B of \citet{riols18c}, in particular the terminal velocity approximation,  it is possible to recast Eq.~(\ref{mass_eq_dust2}) in the useful form of an advection-diffusion equation:
\begin{equation}
\label{eq_dubrulle}
\dfrac{\partial\overline{\rho_d} }{\partial t} = \dfrac{\partial}{\partial z} \left[\left(z \Omega^2+\dfrac{d \Phi_s}{dz}\right)\, \overline{\tau_s} \,\overline{\rho_d} \right) +\dfrac{\partial}{\partial z}  \left[ D_z \, \overline{\rho}  \dfrac{\partial}{\partial z} \left(\dfrac{\overline{\rho_d}}{\overline{\rho}}\right) \right],
\end{equation}
where $D_z \simeq  \langle \overline{v_z^2} \rangle $ $\tau_{\text{corr}}>0$ is  the diffusion coefficient, with $\tau_{\text{corr}}$ the correlation time of the turbulent eddies. Note that the horizontally averaged Stokes number ${\overline{\tau_s}=\text{St} \Omega^{-1} \overline{\rho_0/\rho}}$ is slightly different  from ${\text{St} \Omega^{-1} \rho_0/\overline{\rho}}$.  Due to gas density fluctuations associated with GI, there is a factor $f_c \simeq 1.3 $ difference between the two quantities. This factor is obtained from simulations by averaging in $x$ and $y$ the inverse of the gas density. 
We finally  assume that the gas density can be modelled by a Gaussian ${\rho=\rho_0 \exp{(-\frac{z^2}{2H_g^2})}}$ with midplane $\rho_0= 1.57$ and $H_g=0.44 H$. This hypothesis is not too far from reality for $Q \gtrsim 1$.  Under this assumption, we approximate  ${d \Phi_s}/{dz}\simeq s z \Omega^{2}$  with $s=2.77$ in the limit $z \ll H_g$. 

The equilibrium solution of Eq.~(\ref{eq_dubrulle}) is
\begin{equation}
\label{sol_dubrulle}
\rho_d(z)=\rho_{d_0} \exp\left(-\dfrac{z^2}{2H_g^2}\right)\exp\left(-\int \dfrac{\text{St}\,\Omega f_c (s+1) \,z\,e^\frac{z^2}{2H_g^2}}{D_z(z)} dz \right).
\end{equation}
For uniform diffusion coefficient $D_z$ and $z \ll H_g$,  this gives: 
\begin{equation}
\label{eq_Hdgauss}
\rho_d(z)\simeq \rho_{d_0} \exp\left(- \dfrac{z^2}{2 H_d^2} \right) 
\end{equation}
\begin{equation}
\label{eq_Hdgauss}
 \text{with } \quad \dfrac{H_d}{H_g} =\left(1+\dfrac{\text{St}\,\Omega f_c (s+1)  \,H_g^2}{D_z }\right)^{-1/2}.
\end{equation}
The distribution is Gaussian and the dust scaleheight tends towards unity in the limit of small St. For larger values (but potentially 
still $<1$), the ratio may exhibit the scaling $\text{St}^{-1/2}$.

\section{Simulation without back reaction}
\label{appendixB}

\begin{figure}
\centering
\includegraphics[width=\columnwidth]{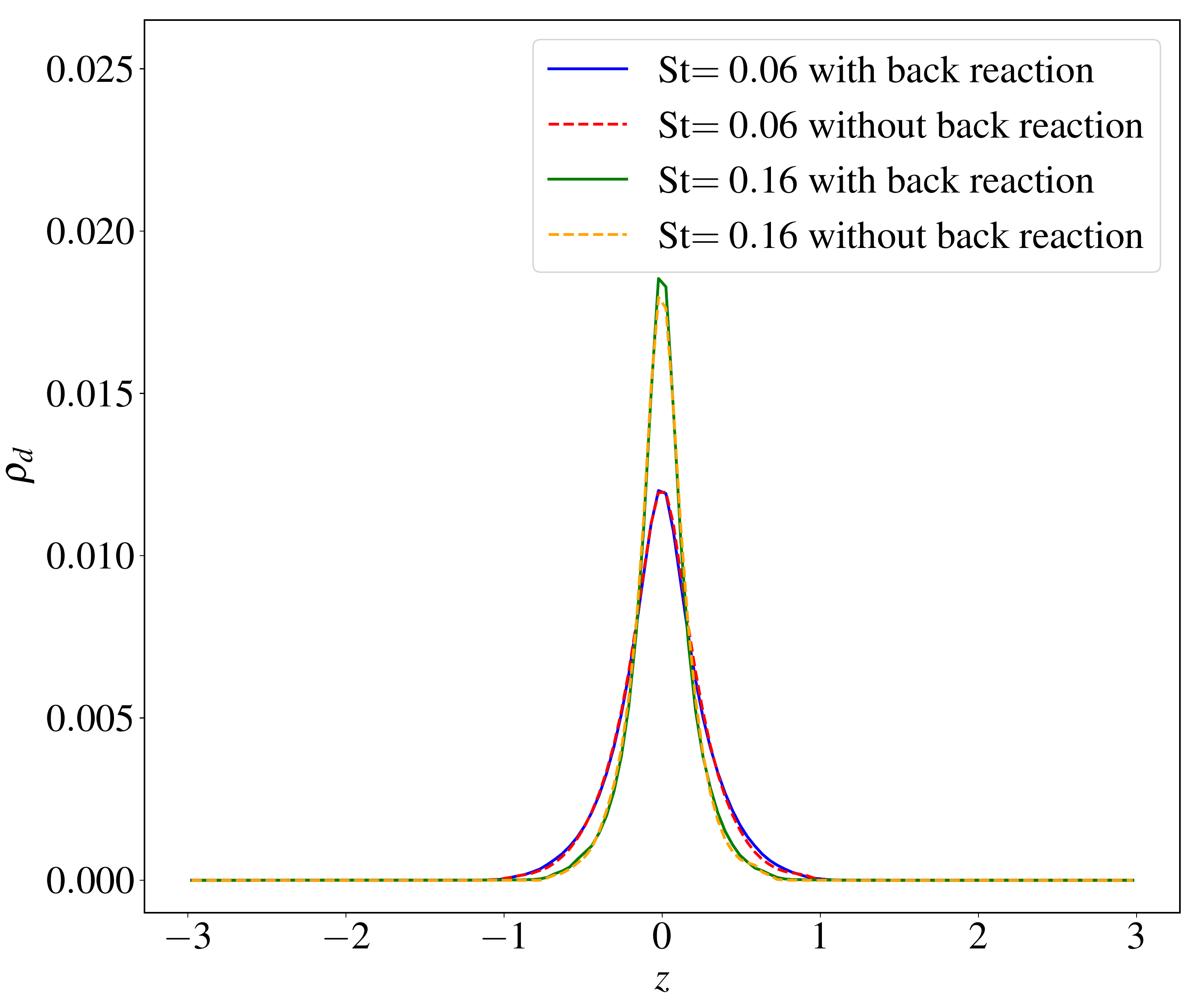}
 \caption{Dust density profiles (averaged in time, $x$ and $y$) for two different Stokes number, with and without back reaction.}
\label{fig_nobr1}
\end{figure}

{We show in this appendix the results of a simulation without the dust back reaction onto the gas. This simulation has been run for $45 \Omega^{-1}$ with resolution of 26 points per $H$ in the horizontal direction and contains two species with $\text{St}=0.06$ and $\text{St}=0.16$. We aim to compare this with other simulations including the back reaction (and same setup and Stokes numbers).  First we show in Fig.~\ref{fig_nobr1} the dust density profile (averaged in $t$, $x$, and $y$). For both Stokes numbers, we find that the profiles are almost indiscernible from each other. This means that the settling process is unaffected by the dust back-reaction.}

\begin{figure}
\centering
\includegraphics[width=\columnwidth]{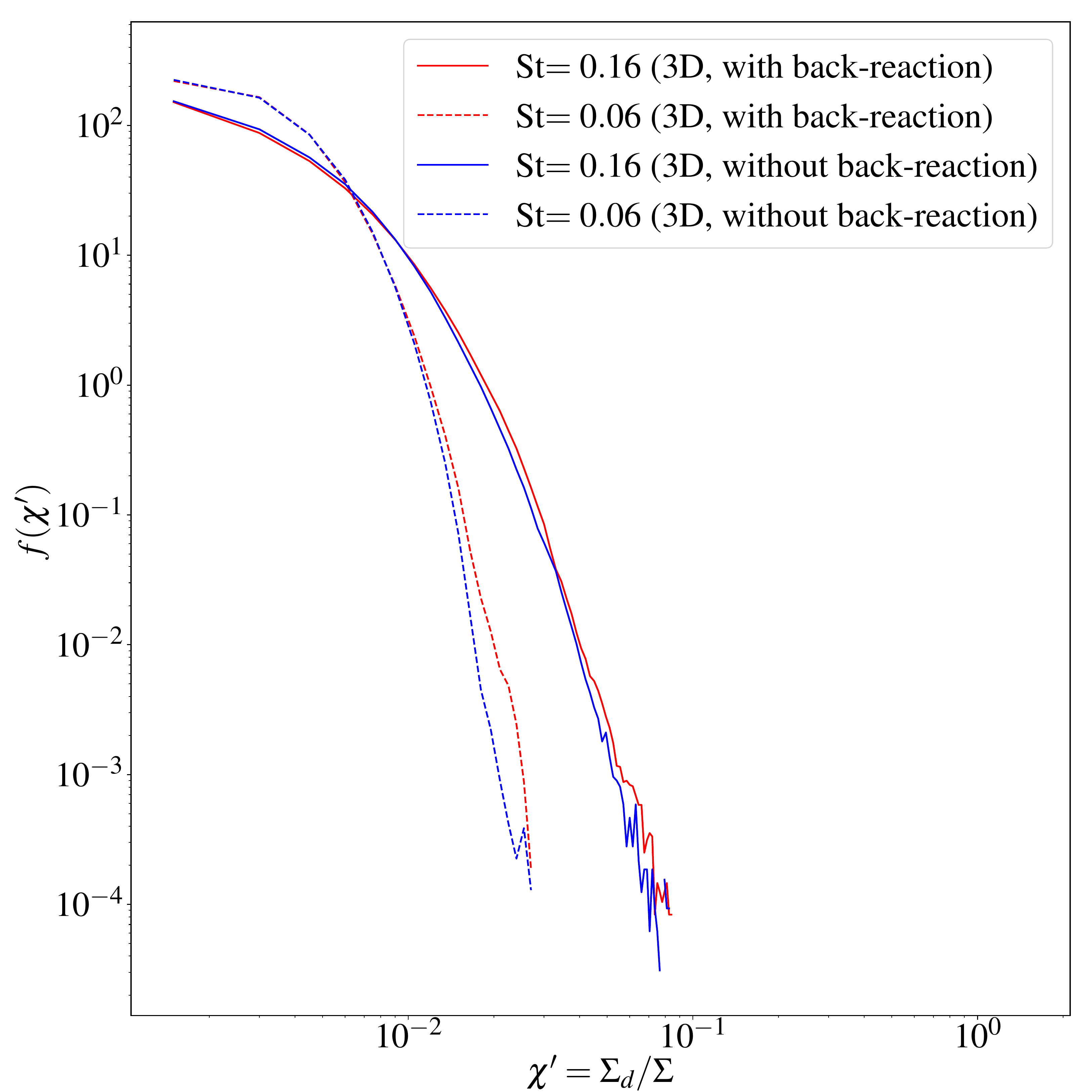}
 \caption{Probability distribution function for concentration events as a function of $\chi'=\Sigma_d/\Sigma$. The dotted lines are for $\text{St}=0.06$ and the solid lines are for $\text{St}=0.16$. We compare the cases with (red) and without (blue) dust back reaction  }
\label{fig_nobr2}
\end{figure}

{To go further, we show in Fig.~\ref{fig_nobr2} the probability distribution function $f(\chi')$ of concentration events (see Section \ref{concentration_events} for details about its calculation), computed for our two different Stokes numbers, in the case with and without back reaction. Again we see only marginal differences between the two cases suggesting that the dust back reaction does not interfere too much with the process of dust concentration and clumping.  A slight deviation is however seen at large $\chi'=\Sigma_d/\Sigma$  for $\text{St}=0.06$ (in the tail of the distribution) but this is expected since the number of events is rare and the statistics not very good at large concentration $\chi'$ (given the time of the simulation). }

\label{lastpage}
\end{document}